\documentclass[sigconf]{acmart}

\usepackage{calc} 
\usepackage{etoolbox}
\makeatletter
\patchcmd\longtable{\par}{\if@noskipsec\mbox{}\fi\par}{}{}
\makeatother
\usepackage{graphicx}
\makeatletter

\usepackage{booktabs}
\usepackage{array}
\usepackage{multirow}
\usepackage{wrapfig}
\usepackage{float}
\usepackage{colortbl}
\usepackage{pdflscape}
\usepackage{tabu}
\usepackage{threeparttable}
\usepackage{threeparttablex}
\usepackage[normalem]{ulem}
\usepackage{makecell}
\usepackage{xcolor}
\definecolor{mypink}{RGB}{219, 48, 122}
\makeatletter

\providecommand{\tightlist}{%
  \setlength{\itemsep}{0pt}\setlength{\parskip}{0pt}}

\copyrightyear{2025}
\acmYear{2025}
\setcopyright{cc}
\setcctype{by-nc-nd}
\acmConference[FSE Companion '25]{33rd ACM International Conference on the Foundations of Software Engineering}{June 23--28, 2025}{Trondheim, Norway}
\acmBooktitle{33rd ACM International Conference on the Foundations of Software Engineering (FSE Companion '25), June 23--28, 2025, Trondheim, Norway}\acmDOI{10.1145/3696630.3728696}
\acmISBN{979-8-4007-1276-0/2025/06}

\title{The Impact of Team Diversity in Agile Development Education}

\author{Marco Torchiano}
\author{Riccardo Coppola}
\author{Antonio Vetrò}
\email{ {first.last}@polito.it}

\affiliation{%
  \institution{Department of Control and Computer Engineering} 
\textit{Politecnico di Torino}
  \city{Turin}
  \country{Italy}
}

\author{Xhoi Musaj}
\email{xhoi.musaj@edu.unito.it}
\affiliation{%
  \institution{Department of Psychology} 
\textit{University of Turin}
  \city{Turin}
  \country{Italy}
}

\begin{document}


\begin{abstract}

Software Engineering is mostly a male-dominated sector, where gender diversity is a key feature for improving equality of opportunities, productivity, and innovation. Other diversity aspects, including but not limited to nationality and ethnicity, are often understudied.

In this work we aim to assess the impact of team diversity, focusing mainly on gender and nationality, in the context of an agile software development project-based course. 

We analyzed 51 teams over three academic years, measuring three different Diversity indexes -- regarding Gender, Nationality and their co-presence -- to examine how different aspects of diversity impact the quality of team project outcomes. 

Statistical analysis revealed a moderate, statistically significant correlation between gender diversity and project success, aligning with existing literature. Diversity in nationality showed a negative but negligible effect on project results, indicating that promoting these aspects does not harm students' performance. Analyzing their co-presence within a team, gender and nationality combined had a negative impact, likely due to increased communication barriers and differing cultural norms.

This study underscores the importance of considering multiple diversity dimensions and their interactions in educational settings. Our findings, overall, show that promoting diversity in teams does not negatively impact their performance and achievement of educational goals.
\end{abstract}


  
\maketitle

\section{Introduction}\label{introduction}

The lack of diversity in the tech industry is a well-known issue \cite{Chakravorti20}, even though several studies have recognized perceived diversity (i.e., the diversity factors that individuals are born with) as a high-value team property within Software Engineering (SE) teams, with companies willing to increase their effort to create more diverse work teams \cite{rodriguez2021perceived}. 

Regarding gender diversity, many studies have demonstrated that it brings benefits to the SE workplace, e.g. with increases in innovation \cite{vasilescu2015gender}, productivity \cite{Ortu2017}, and with higher sentiment, although inequities are still evident: Guzmán et al. \cite{Guzmn2023}  reported that while both women and men hold leadership responsibilities in similar numbers, women are more likely to feel unsupported in their roles, have less authority, be dissatisfied with their compensation, and feel undervalued within their teams. Furthermore, their survey highlights that women face greater challenges related to team dynamics and biases, whereas men’s primary barriers are technical and project-related issues.

Evidence about the increasing number of women (and of their responsibilities) collected from workplaces is however in contrast with the current trends in the educational context. In a snapshot presented in \cite{anderson2024learning}, related to Software Engineering education in a UK University, the female student population remains critically low, barely rising above 18\%. 
Anh Nguyen-Duc and Letizia Jaccheri found that the active involvement of female students occurs in project management and requirement engineering, while the areas lacking active involvement are software architecture and Scrum methodology. \cite{NguyenJaccheri}.

Other aspects of diversity appear understudied in the related literature: Szlavi et al. \cite{Szlavi23} underline the necessity of addressing the systemic barriers in CS through intersectional lenses\footnote{The concept of \emph{Intersectionality} is defined by Crenshaw as how race, class, gender, and other individual characteristics “intersect” with one another and overlap \cite{crenshaw2013demarginalizing}.}, and that an encouraging growing awareness can be found in the field. However, many of the provided results exploring these aspects are mixed, e.g. regarding nationality, age, race, disability, and sexual orientation. In the aforementioned analysis by Ortu et al.  -- for instance -- it is indicated that diversity in nationality is linked with lower team politeness \cite{Ortu2017}; another study by Stahl et al. finds that cultural diversity has a negative impact on SE processes \cite{Stahl2010}. Sanchez-Gordon and Colomo-Palacios denounce a long-standing lack of diversity in Computer Science, that can create a toxic culture and contribute to low rates of participation by underrepresented groups (URGs) \cite{Sanchez2024}. 

The main objective of the present study is to perform an analysis of the current status in terms of DEI (Diversity, Equity, and Inclusion) in a Software Engineering capstone course held at Politecnico di Torino. To the best of our knowledge, this paper constitutes the first quantitative investigation about how DEI aspects influence team performance in Software Engineering courses. The course includes a mandatory team project, where the students are tasked with the development of a small-scale software project using the Agile and Scrum methodologies, thus fostering constant collaboration and opinion exchange between team members. All the project teams have been created by allocating the students to maximize the average diversity within each team. In this manuscript, we report our conclusion regarding the effects on course appreciation and team performance, in three consecutive editions of the course.

The remainder of this paper is structured as follows: in Section II, we provide a survey of prominent results about diversity and inclusion in Software Engineering; in Section III, we discuss the methodology used for the evaluation and the context where it took place; in Section IV, we present the results, that are discussed in Section V, along with the main threats and current limitations of the study; finally, Section VI concludes the paper and provides possible future research directions. 

\section{Background}\label{background}

In this section, we report findings from related works in the literature analyzing the effects of diversity within SE teams, and the metrics that can be used to quantitatively measure such diversity.

\subsection{Effects of team diversity in SE}\label{team-diversity}

Many studies in the recent literature have analyzed, at different
education levels and in different countries, how diversity and inclusion
topics are perceived by SE teachers and students, and which measures are
taken to approach the topic.

Happe et al.~presented a study on the gender gap in SE education and
careers in Germany, through a retrospective analysis of women's
experiences at the university level. The author's findings reveal that
significant barriers -- such as the perception of software engineering
as incompatible with personal interest, a male-dominated environment,
and a lack of female role models -- deter many women from pursuing the
field. Persistent stereotypes and social challenges contribute to gender
inequality, underscoring the need for supportive measures to encourage
diversity in computing education \cite{happe2024decoding}.

Anderson et al.~examined gender disparities in SE education within a UK
university, focusing on both undergraduate and postgraduate levels. The
study analyzes institutional data, national benchmarks and a survey of
Master's students. The primary findings reveal a persistent gender
imbalance, with women significantly underrepresented, particularly at
higher academic levels. The authors point out that female students at
the studied institution achieve comparable academic performance to their
male counterparts, indicating no significant attainment gap, though
female enrollment remains low. The authors underline how these results
mirror broader UK trends and the need for interventions to generally
support female participation in SE \cite{anderson2024learning}.

Cutrupi et al.~presented the Draw-A-Software-Engineer Test (DASET), a
research tool to study the perception of societal norms, gender biases
and stereotypes. The authors asked nine university students to draw and
describe their perception of software engineering. Of the artefacts,
78\% of the drawings represented a man as a software engineer.
Additionally, only 11\% of the drawings included a software engineer
with brown skin \cite{cutrupi2024draw}.

Hyrynsalmi studied the approaches of diversity and inclusion most
commonly used by SE teachers in Finland, and how the importance of
diversity and inclusion is perceived in their courses. The author found
that whilst some of the respondents were hesitant to introduce such
aspects, the importance of diversity and inclusion in SE education was
recognized by most. It is underlined an existing familiarity with
accessibility approaches, that are often used as an example of diversity
and inclusion in teaching. Finally, the respondents recognized an
increase in diversity and multiculturalism in the academic field, and
the presence of trust issues between individuals that prioritize
diversity and inclusion topics and those who do not
\cite{hyrynsalmi2023diversity}. Hyrynsalmi et al.~also approached the
topic of diversity in SE paths through an online survey sent to their
students, showing that on average women apply later to SE studies than
men, with statistically significant differences between genders
\cite{hyrynsalmi2024second}.

Kovaleva et al.~pointed out the necessity for the design of gender-neutral
Software Engineering programs, as a tool to reach gender balance in the
domain. The authors underlined the necessity of adopting a gender-neutral
language in marketing and communication for SE programs, pointing out
the existing pressure of stereotypes experienced by both genders
\cite{kovaleva2022designing}.

Several studies have also analyzed diversity and inclusion for SE
practitioners. Petrescu et al.~conducted a study in Romania to
investigate gender diversity in the Scrum Master role, within the
context of agile software development, focusing on professionals in
various companies. Primary findings suggest that women are well-suited
to the Scrum Master role due to their strong communication, empathy, and
conflict-management skills, valued more than technical
expertise in this role. These findings at the same time may however
underline a trend to underestimate women's hard skills in favor of soft
skills \cite{petrescu2023women}.

Zhao and Young analyzed workplace discrimination in Software Engineering
through a questionnaire administered to 97 respondents from all over the
world. The findings highlighted that the main causes of workplace
discrimination in SE are age, gender, and race/ethnicity/nationality.
The intersectionality of the discrimination issue is evident when the
authors point out that when analyzing other demographic factors than
gender, female software professionals are more likely to be
discriminated \cite{zhao2023workplace}.

\subsection{Diversity indexes}\label{sec-di}

For our analysis of diversity within teams in SE, we considered all the three diversity indexes that were mentioned in the analyzed related work.

  \textbf{Richness}: the number of different types present in the population.

  \textbf{Shannon index} \cite{Shannon}: originally proposed in information
  theory, the Shannon entropy quantifies the uncertainty in predicting
  the type of individual that is taken at random from the dataset.
  It is defined as:
  
  $$H = \sum_{i=1}^{R} p_i \cdot ln(p_i)$$
  
  where $R$ is richness (the
  total number of types in the dataset), $p_i$ is the proportional
  abundances the $i$-th type.

  \textbf{Blau diversity index} \cite{blau1977}: The measure equals the
  probability that two entities taken at random from the dataset of
  interest represent the same type. It is defined as:

  $$B = 1 - \sum_{i=1}^{R} p_i^2$$
  

In table ~\ref{tbl-DI-examples}, as an example, we show the corresponding diversity values for a population of 12 individuals with different random exemplar compositions.

\begin{table}

\caption{\label{tbl-DI-examples}Diversity Indexes for
exemplar team compositions (letters symbolize different types)}

\centering{
\footnotesize
\begin{tabular}[tb]{lrrr}
\toprule
\textbf{Data} & \textbf{Richness} & \textbf{Blau} & \textbf{Shannon}\\
\midrule
\texttt{A b b b b b b b b b b b} & 2 & 0.153 & 0.287\\
\texttt{A A A A A A b b b b b b} & 2 & 0.500 & 0.693\\
\texttt{A A A b b b C C C C C C} & 3 & 0.625 & 1.040\\
\texttt{A A A A b b b b C C C C} & 3 & 0.667 & 1.099\\
\texttt{A b C C d d d d d d d d} & 4 & 0.514 & 0.983\\
\texttt{A A A b b b C C C d d d} & 4 & 0.750 & 1.386\\
\bottomrule
\end{tabular}

}

\end{table}%

\begin{figure*}[bt]
    \centering
    \includegraphics[width=1\linewidth]{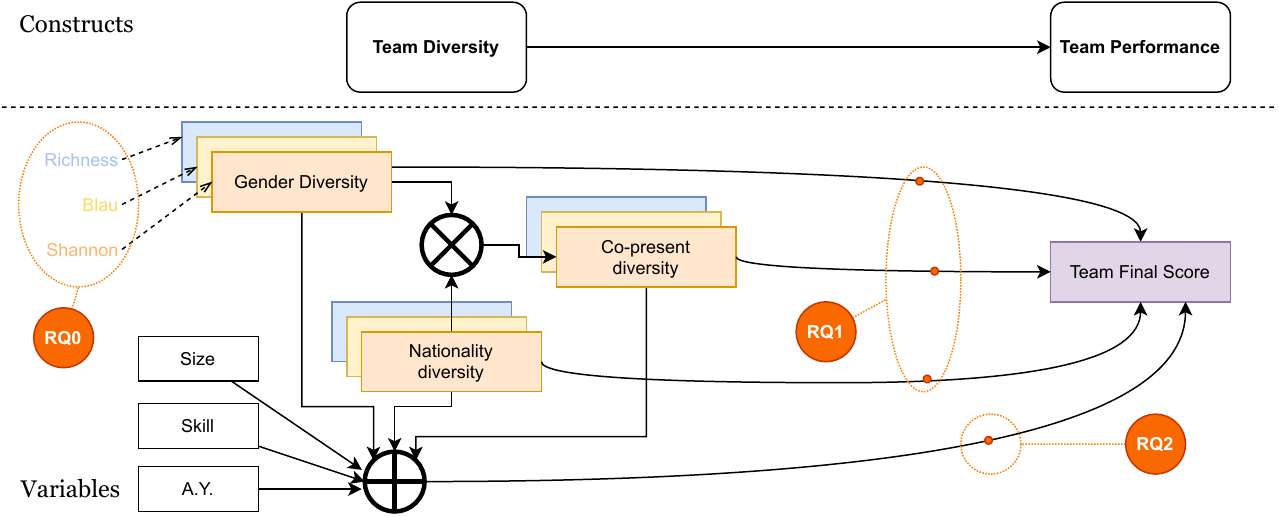}
    \caption{Overview of study design.}
    \label{fig:study-design}
    \Description{Schematic representation of study design,     including constructs and the relative variables,     with links among them and indication of RQs}
\end{figure*}

\section{Method}\label{method}

The general overview of the design of our study is illustrated in figure \ref{fig:study-design}.
This section details the context of our analyses, the research questions, the variables used to answer them, and the analysis method applied.

\subsection{Context}\label{context}

The setting for the study reported in this paper is Master Software Engineering course. The main educational goal of the course is to provide the students with
the essential skills and knowledge required to manage and execute
software projects using agile methodologies. The focus is on instilling
a comprehensive understanding of the agile development approach,
emphasizing hands-on experience in carrying out team-based software
projects efficiently and in an inclusive manner, i.e.,~enhancing the
peculiarities of each person. All of the students involved have extensive experience in teamwork, acquired in previous courses; moreover, the instructors constantly monitor the behavior of the teams to capture possible communication issues and provide feedback for improvement.

Teams are formed after the first week, based on the information declared by the students in the preliminary questionnaire. We aim to achieve homogeneity
between teams by balancing a variety of aspects that are assessed
through an online form; the main aspects are gender, nationality,
technical skills, and previously attended courses.

The actual composition of teams may vary due to some students dropping
out of the course since they are not able to regularly attend the classroom practical activities and participate in the project work.

From the Gender Balance Report of our university \cite{polito2024}, we know that
overall female students are 30.5\%, while in computer-related degrees
the female students are 16.8\% (in line with the findings reported in
\cite{anderson2024learning}). Focusing on nationality diversity, on average the
university enrolls 20.6\% foreign students.

The grading is performed by evaluating both team and individual
performance. Team evaluation weights 80\%, while individual evaluation
weights 20\%. 
The team evaluation is carried on according to four main
criteria: team coordination and improvement, Scrum compliance, quality of meetings, and doneness.


\subsection{Research questions}\label{research-questions}

Since the novelty of this kind of studies especially in the SE domain, we defined a preliminary research question focused on methodology:

\begin{itemize}
\item[\textbf{RQ0}:] \textit{How are different diversity indexes related to each other?}

  Among the several diversity indexes proposed in the literature, we picked the 
  three most common -- presented in Sec.~\ref{sec-di} above --. 
  We are interested in understanding how they are related to
  each other and if there is a preferable one.

\end{itemize}

The main goal of our investigation is to understand how team diversity plays
a role in agile software development projects and specifically how it can affect team performance. To this end, we formulate two main research questions:

\begin{itemize}
\item[\textbf{RQ1}:]
  \textit{What is the impact of diversity factors on team performance?}

  We consider different team diversity factors, measured according to
  diverse indexes, and observe the correlation with the team score.

\item[\textbf{RQ2}:] \textit{What is the relative importance of different diversity factors?}

  We observe the magnitude of the impact of different diversity factors
  to understand which ones play a more meaningful role.
\end{itemize}

\subsection{Variables}\label{variables}

In order to address the main research questions, we consider a single dependent construct which is team performance. We operationalise it through one variable which is the final score assigned to the team. 
In particular, since the top grade differed in the three considered academic years, we adopt a normalized version so that 100 is the top grade attributable for the specific academic year.

As far as the factors whose influence we wish to study (i.e., the independent variables), they are:

\begin{itemize}
\tightlist
\item
  the \textbf{Gender diversity}, measured on the basis of the self-declared
  gender of the student, according to the definition\footnote{Gender (n): the condition of
being male, female, or neuter. In
a human context, the distinction
between gender and sex reflects
the usage of these terms: Sex
usually refers to the biological
aspects of maleness or
femaleness, whereas gender
implies the psychological,
behavioural, social, and cultural
aspects of being male or female
(i.e., masculinity or femininity) } and guidelines provided by the American Psychological Association \cite{american2015guidelines}.
\item
  the \textbf{Nationality diversity}, measured on the basis of the official
  nationality declared by the student at enrollment,
\item
  the \textbf{Co-present (Gender and Nationality)} diversity, measured as the
  joint class of Gender and Nationality, 
\item
  the \textbf{Career diversity}, measured on the basis of the student being an Erasmus or International student, i.e.~students who do not share a common background
  with the other students, since they typically come from foreign university for a single semester.
\end{itemize}

All diversity factors have been measured using the three diversity
indexes presented in section Section~\ref{sec-di} above. Therefore we
have nine different combinations of factor and diversity index.

In addition, we also monitored a few co-factors that could affect the
final team score:

\begin{itemize}
\tightlist
\item
  the \textbf{size} of the team,
\item
  the \textbf{skill score}, measured as self reported skill in basic
  software development, web development, and software engineering (collected through a questionnaire administered at the beginning of the course),
\item
  the \textbf{academic year} of attendance.
\end{itemize}

\subsection{Analysis method}\label{analysis-method}

Concerning RQ0, we report the distribution of three different indexes for the considered diversity factors. In addition, we compute the overall correlations among the indexes.

Concerning RQ1, we compute the correlation coefficients as well as the relative statistical significance. Since we expect distributions far from normal we opted for the Spearman correlation coefficients which is
suitable for non-normal distributed data.

As far as RQ2 is concerned, we compute a linear regression model of the final team score vs.~all the diversity factors and the co-factors that we
monitored.

Overall the goal of the study presented in this paper is exploratory, we do apply statistical tests as guidelines. In deciding whether to decide statistical significance we e consider the customary $\alpha = 0.01$ as the threshold.

We aim to interpret the results and provide a general assessment of the effects of diversity on the team performance.




\section{Results}\label{results}

In this section, we report information about the population of students evaluated in our study and the results of the three research questions described in the previous section.

\subsection{Population}\label{population}

We consider the teams that participated in the latest three completed
edition of the course, i.e. in academic years from 2021/22 to 2023/24.

Overall, we analyzed 51 teams over three academic years, which included a total of 324 members.

Table~\ref{tbl-team-summary} reports the summary statistics of
team compositions in the three academic years we considered in our
analysis: for each year we report the number of teams, the distribution of team size, the average percentage of women in teams, the average percentage of foreign students in teams, and the average grade obtained by the team at the end of the academic year.

\begin{table}

\caption{\label{tbl-team-summary}Summary of team compositions}
\footnotesize
\centering{

\begin{tabular}[t]{@{}l@{}rr@{~}r@{~}rll@{}r@{}}
\toprule
&& \multicolumn{3}{c}{Size} & & & Average\\
\cmidrule{3-5}
AY & Teams & Min & Median & Max & Women \% & Foreign \% & Score\\
\midrule
2021/22 & 13 & 5 & 7 & 7 & 15.3\% & 14.1\% & 93\\
2022/23 & 20 & 5 & 6 & 7 & 16.7\% & 25.8\% & 91\\
2023/24 & 18 & 5 & 7 & 7 & 18.5\% & 27.7\% & 90\\
\bottomrule
\end{tabular}


}

\end{table}%

We observe that the size of teams is generally 6 or 7 members. We
originally formed teams of the same size based on the online
questionnaire that the prospective students filled in before the start of the
course (see also section \ref{context}), but eventually few
teams ended up with fewer components due to some students not showing up at the beginning of the course lectures or dropping out after a few lectures. The percentage of women increased
from 15\% to 18\% in the three years, while the proportion of foreign
students doubled (from 14\% to 28\%).

\begin{figure}

\centering{

\includegraphics[width=0.95\columnwidth]{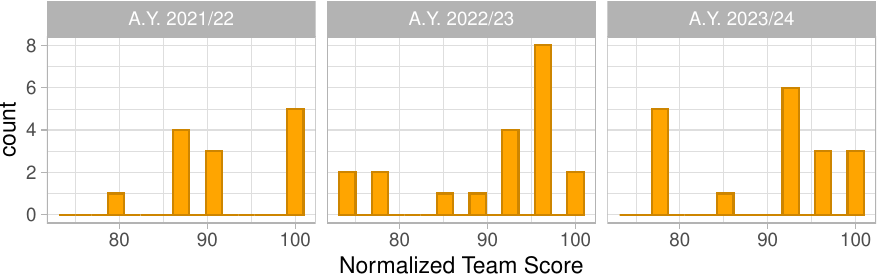}
}

\caption{\label{fig-team-score}Team score distribution}
\Description{Distribution of team scores for the three academic years}

\end{figure}

The average normalized team scores ranged from 93 to 90. Figure~\ref{fig-team-score} reports the detailed distribution of team scores in the three considered academic years.

\subsection{RQ0 - Relationship among
\label{rq0}
indexes}\label{rq0---relationship-among-indexes}

\begin{figure}
    \centering
    \includegraphics[width=0.95\columnwidth]{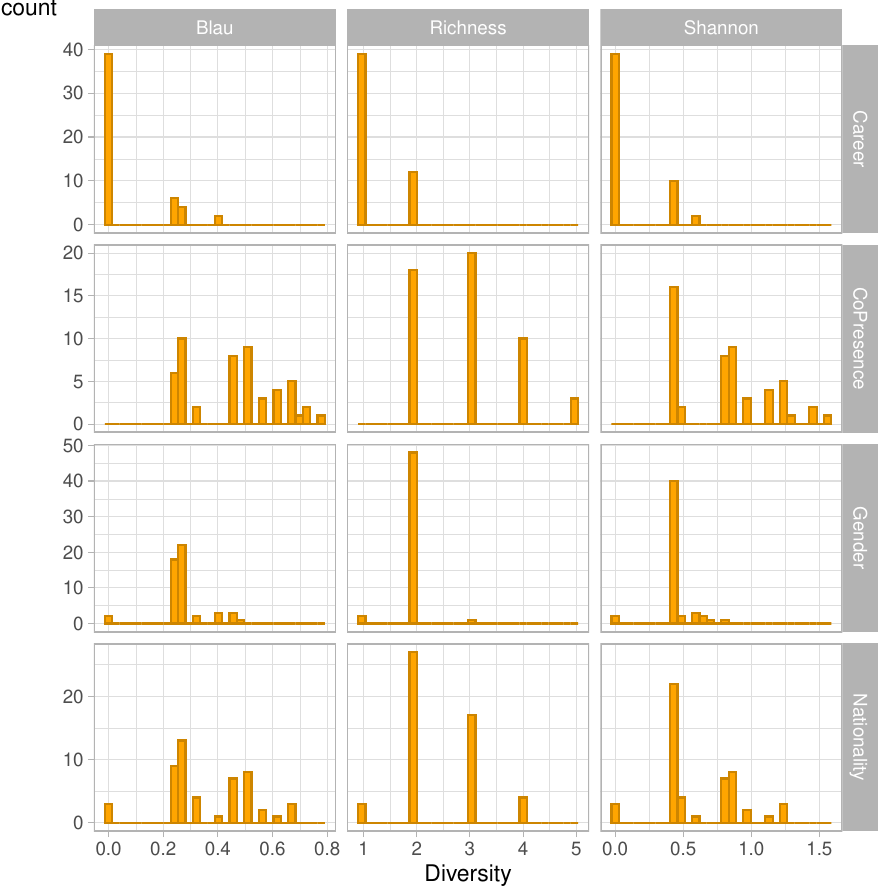}
    \caption{\label{fig-di-distributions}Overview of DI distributions}
    \Description{Distribution of all combinations of diversity features and relative measurement index}
    \end{figure}

The distribution of the values of the three Diversity Index applied to each of the four Diversity Factors is reported in
Figure~\ref{fig-di-distributions} that shows $3 \times 4 = 12$ histograms.

We observe that Richness (mid column) includes a few distinct
values, as we would expect from its definition. While Blau and Shannon indexes (the two outer columns) are more spread over different values.
When considering the indexes -- i.e. looking by row -- we can observe similar patterns.

\begin{figure}
\centering{
\includegraphics[width=0.90\linewidth]{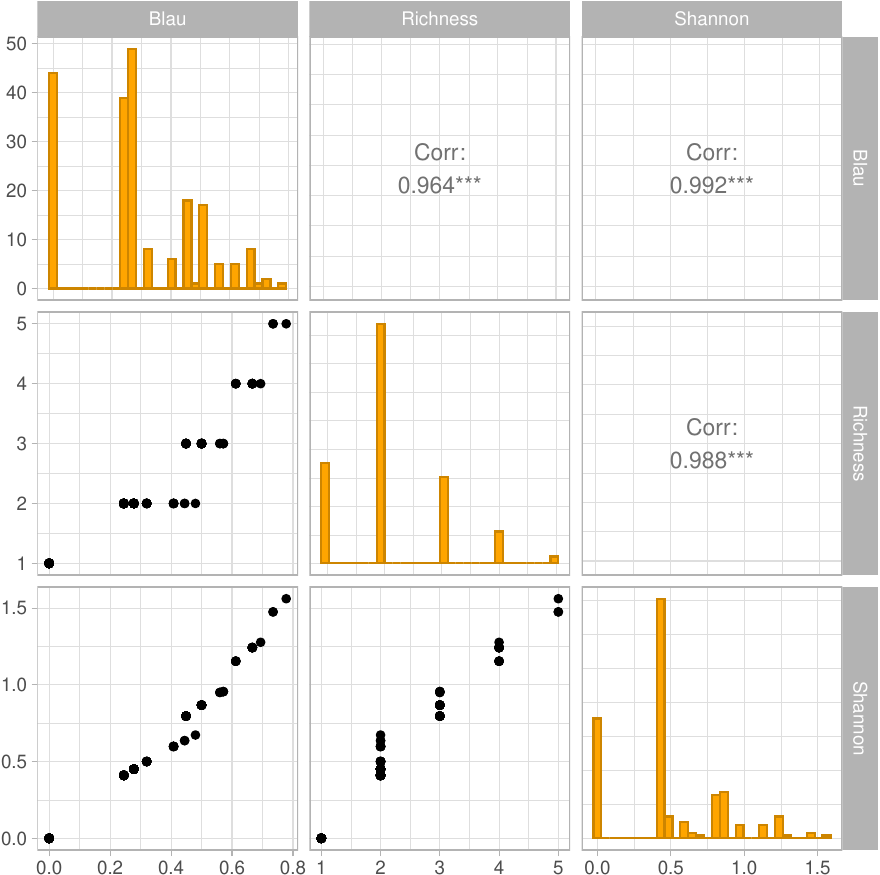}
}

\caption{\label{fig-DI-cross}Cross correlations of the three Diversity Indexes}
\Description{Cross correlations of the three Diversity Indexes: reports pair scatterplots, distribution histograms, and correlation values which are all above 0.96}
\end{figure}

A more holistic view is reported in figure~\ref{fig-DI-cross}, where for each index the values computed across all factors are considered together. In practice we summed together the histograms in the three columns of the previous figure, ignoring the distinction among the factors.
In particular we report the correlation among the three different indexes.

There is a generally high correlation between the three pairs of Indexes, the highest being between Blau and Shannon. Since the Richness index is -- by construction -- discrete, it shows a slightly lower correlation with the other two indexes.

We also observe in the scatter plot of Blau vs.~Shannon and Blau
vs.~Richness (two leftmost) that for a few values the two pairs of indexes are inconsistent. 
In particular, the ordering of two specific configurations is different w.r.t Blau vs.~Shannon and Richness. The
inconsistent pair corresponds to a configuration of Gender in a 5-member team and a configuration of Nationality in a 7-member team, whose details and index values are reported in table~\ref{tbl-inconsistent}.
Since such inconsistency involves two different features it does not
affect the remainder of our analysis.
However it must be noted if we aim to apply diversity indexes in larger data sets.

\begin{table}

\caption{\label{tbl-inconsistent}Configuration causing index
inconsistencies}

\centering{
\footnotesize
\begin{tabular}[t]{llrrr}
\toprule
\textbf{Feature} & \textbf{Configuration} & \textbf{Blau} & \textbf{Shannon} & \textbf{Richness}\\
\midrule
Nationality & IR IT IT IT IT IT PT & 0.449 & 0.796 & 3\\
Gender & F F M M M & 0.480 & 0.673 & 2\\
\bottomrule
\end{tabular}

}

\end{table}%

\subsection{RQ1 - Impact of team
diversity}\label{rq1---impact-of-team-diversity}

\begin{figure}

\centering{

\includegraphics[width=0.95\columnwidth]{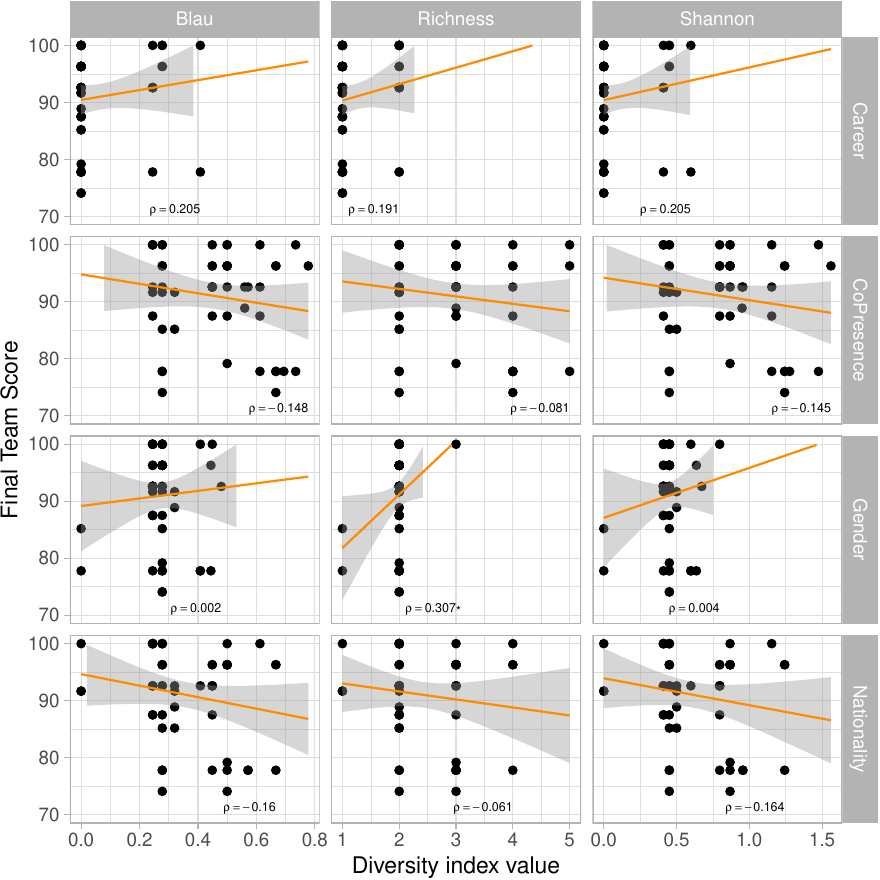}

}

\caption{\label{fig-correlations}Correlations of Diversity and Team
Score for different DI and factors}
\Description{Correlations of Diversity and Team Score for different DI and factors}
\end{figure}

Correlation coefficients of the different diversity measures vs.~the
final team score are reported in table~\ref{tbl-correlations}.
Figure~\ref{fig-correlations} reports the scatter plots of the
normalized final team score vs.~the different combinations of diversity
indexes and factors. The figure also reports the linear regression line
together with the confidence interval and the value of the Pearson
\(\rho\) coefficients.

\begin{table}

\caption{\label{tbl-correlations}Correlations of Individual Diversity Indexes values and Team Score}

\centering{
\begin{tabular}[t]{@{}llrr@{}l@{}}
\toprule
\textbf{Feature} & \textbf{Index} & \textbf{Spearman $\rho$} & \textbf{p-value} & \\
\midrule
Career & Blau & 0.205 & 0.149 & \\
Career & Richness & 0.191 & 0.179 & \\
Career & Shannon & 0.205 & 0.149 & \\
\addlinespace
CoPresence & Blau & -0.148 & 0.300 & \\
CoPresence & Richness & -0.081 & 0.573 & \\
CoPresence & Shannon & -0.145 & 0.309 & \\
\addlinespace
Gender & Blau & 0.002 & 0.989 & \\
Gender & Richness & 0.307 & \textbf{0.028} & *\\
Gender & Shannon & 0.004 & 0.978 & \\
\addlinespace
Nationality & Blau & -0.160 & 0.263 & \\
Nationality & Richness & -0.061 & 0.672 & \\
Nationality & Shannon & -0.164 & 0.251 & \\
\bottomrule
\end{tabular}

}

\end{table}%

We observe that almost all the correlations are not statistically
significant. The only exceptions is the Gender Richness index (p=0.028) whose correlation index is $\rho=0.307$, that can be considered a moderate strength correlation.

We also note that, although not significant, all correlations
those involving Gender and Career are positive.
While Nationality and Co-Presence are negative.

\subsection{RQ2 - Combined Relative
Importance}\label{q2---combined-relative-importance}

Finally, we evaluate the combined influence of the different diversity factors, as well as the additional co-factors that we monitored, using a linear model. We selected one index for each diversity feature based on the strengths of their individual correlations. 
Table~\ref{tbl-linear-model} reports the estimated values of the
coefficients for all the factors. In addition, it reports the results of an ANOVA test.

\begin{table}

\caption{\label{tbl-linear-model}Comprehensive linear model of diversity
factors and additional co-factors.}

\centering{


\begin{tabular}[t]{@{}l@{}rrrr@{}}
\toprule
\textbf{Coefficient} & \textbf{Estimate} & \textbf{Std. Error} & \textbf{t value} & \textbf{Pr(>|t|)}\\
\midrule
(Intercept) & 32.327 & 37.535 & 0.861 & 0.394\\
Skill & 38.888 & 35.541 & 1.094 & 0.280\\
Gender Richness & 11.185 & 5.248 & 2.131 & \textbf{0.039}\\
Career Shannon & 5.918 & 6.777 & 0.873 & 0.388\\
Nationality Shannon & 2.910 & 7.071 & 0.412 & 0.683\\
\addlinespace
CoPresence Shannon & -7.219 & 5.733 & -1.259 & 0.215\\
size & 1.614 & 2.282 & 0.707 & 0.483\\
AY2022\_23 & 4.888 & 4.850 & 1.008 & 0.319\\
AY2023\_24 & -1.419 & 3.514 & -0.404 & 0.688\\
\bottomrule
\end{tabular}

}

\end{table}%

From the overall model, we have the confirmation of only one factor for which we observed a significant correlation, that is the Gender measured
with the Richness index. We use the model to understand how different diversity factors interplay with each other and with the co-factors.

We can roughly compare the different coefficients to each other since the different diversity indexes as well as the Skill have a similar range. Considering all the considered factors and co-factors we observe
that the largest coefficient is linked to Skill which accounts for roughly 40\% of the team grade (expressed in a 0-100 range). The second largest coefficient corresponds to Gender diversity, which
positively affects team performance. Career diversity and Nationality diversity have a small positive effect on the team performance (5.92 and 2.91). The only factor having a negative absolute value for the estimate (-7.22) is CoPresence  -- i.e.~the combined mix of Gender and Nationality within a team.

In addition, we observe that the contextual factor academic year influences the team grade, and that team size has a positive effect on the team performance, even though we cannot compare it directly to the other indexes since it has a value ranging from 5 to 7.

\section{Discussion}\label{discussion}

The findings presented in the previous section allow us to address the three research questions.

\subsection{Answers to Research Questions}

We observed a general consistency and high correlation among the
considered diversity indexes (RQ0). Nevertheless, we found a small
inconsistency of order. While it can be considered practically
irrelevant for our study, it represents a reminder that even the apparently simple question about what is more or less diverse does not have a unique answer. In general, we emphasize that critical adoption of numerical indexes carries the risk of hiding complexity behind the surface of presumed objectivity.

The main objective of this study was an investigation of the effects of different diversity aspects on the quality of the project works performed by teams in a Software Engineering course focused on agile software development (RQ1). Through statistical analysis, we found that gender diversity has a moderate-sized correlation with the final outcome of the project (statistically significant). This outcome is in agreement with related studies in the literature. 

The Nationality Diversity aspect had instead a negative but negligible (non-significant p-value, very far from the 0.05 selected threshold) effect on the results of the team projects. The non-significance of the results is deemed as an important result in demonstrating that promoting nationality diversity within groups has no relevant negative effects on the final results and grades obtained by students in a master's course. Thus, the diversity aspect shall not be sacrificed in favour of a tendency of students to form teams of a single nationality, which we still witness when teams are not defined by the teachers.

In addition to the effect of nationality on the project results, we have also evaluated the effects of the \emph{Career} factor, regarding the fact that a foreign student was currently in an Erasmus programme (i.e., for a semester). Also in this case, we found a positive yet non-significant effect. Even though finer analysis would be required to investigate this phenomenon, we speculate that this aspect might be related to some pre-selection aspects inherent in the Erasmus programme selection procedures, i.e., there might be a dependency between the Career diversity and the Skill diversity.

Finally, the CoPresence that represents the intersection of Gender and Nationality has a negative, although not statistically significant effect.



When considering a combined model of the independent variables (RQ2) things change from considering the different factors in isolation. In this case, the effects (positive and negative respectively) of gender is confirmed. By contrast, in this case, the Nationality Diversity has a positive effect on final score. In the combined model, whilst both gender and nationality have a positive effect if considered individually, co-present diversity (the combination of gender and nationality within a team) had instead a negative effect on the group. This leads us to speculate that despite the isolated benefits of gender and nationality diversity, their co-presence within a team can lead to exacerbation of misunderstandings due to higher communication barriers and different cultural norms related to gender roles and interpersonal communication. All these findings are in line with the corpus of research in Global Software Engineering \cite{wu2012overview}.

Overall, as expected, the most impacting factor on the final project results and scores was the skill level of the students, although not statistically significant.
The lack of significance is probably due to an approximate measurement of skill, through a self-reporting questionnaire at the beginning of the course.

As a qualitative experience report from the studied three editions of the course, we witnessed the occurrence of a few issues between team members and scarce integration of foreign students within groups. Specifically, these issues happened for four groups in a.y. 2021-2022, three groups in 2022-2023, and four groups in 2023-2024. Two of these conflicts involved students in the Erasmus programme. All such criticalities were taken into account during meetings with the groups, which were encouraged to provide a more inclusive support to team members by keeping into account possible difficulties in communication or discrepancies in the respective skill sets. All the groups involved proved receptive to the suggestions.
In the same period, we did not witness similar criticalities related to the other considered diversity aspects.

Overall we believe that the negative -- albeit very limited -- consequences of higher diversity mostly stem from the difficulty in collaborating with people coming from a different -- although within the same discipline -- background.
We think that, from an educational perspective, exposing the student to different and challenging situations can have beneficial effects. Especially in preparing them to face circumstances that would likely happen in work settings, but in a relatively safe learning context.


\subsection{Threats to Validity and Limitations}
One potential threat to internal validity in this study is the selection of variables chosen to measure team diversity and its effects. This was controlled by using three different indexes: we found a general consistency and high correlation among the diversity indexes (see \ref{rq0} ), with only minor order inconsistencies.

Other limitations impacting internal validity include: the  short time span of the course (one semester, with the most relevant activities consisting of four two-weeks Scrum prints); the relatively small number of teams observed (51); and borderline statistical significance achieved. 

Concerning  construct validity, the primary threat arises from the data collection methods: evaluation of teams is partially based on self-reporting data on gender and skills by students. 

As far as conclusion validity is concerned, we stress the fact that we adopted a fairly permissive value of $\alpha$ since our is an exploratory study. The low p-value for the correlation of Team Final Score and Gender Richness (0.028) is mainly related to the presence of a single person in the sample who preferred not to disclose their gender, when removing the relative team the resulting p-value (0,099) would still be significant.

Finally, a significant threat to external validity is the limited applicability of the findings to other contexts. The results of this study may not be generalizable beyond the specific course or educational setting in which the research was conducted. Factors such as the unique characteristics of the student population, the specific Agile methodologies used (Scrum), and the cultural context may limit the broader applicability of the findings.

An inherent limitation of the study in its current state is the lack of an analysis of other forms of diversity (e.g., age, sexual orientation, disabilities). For age and disability, the data at hand was incomplete. Regarding disability, we only had a single data point with such diversity which was not sufficient to perform an analysis on that aspect. Additionally, the analysis of the diversity in nationality (i.e., country of origin) can be considered as a proxy but not directly mappable to the ethnic diversity of the students. Future studies may involve the administration of questionnaires to capture the distribution of these diversities over the student sample, in order also to capture the concept of Intersectionality as established in the literature \cite{crenshaw2013demarginalizing}.

\subsection{Comparison with related work}

The work detailed in this paper is in line with several analyses performed in the literature, even though to the best of our knowledge it is the first investigation of DEI topics in advanced Software Engineering courses revolving around the Agile philosophy of development, and the first research item quantitatively assessing the impact of intra-team diversity on team project results. 

Hyrinsalmi et al. \cite{hyrynsalmi2023diversity} investigated how Finnish universities address diversity and inclusion (D\&I) in software engineering education through a survey of 30 faculty members. While the study acknowledges the rising importance of D\&I in industry and academia, it finds that current teaching methods need improvement, particularly in raising awareness among educators. The focus is on identifying common practices, challenges, and opportunities for embedding D\&I into curricula. The emphasis of the work lies on institutional strategies rather than practical, team-based educational activities.

A second work by Hyrinsalmi et al. \cite{hyrynsalmi2024challenges} extends the previous study by presenting a framework for integrating D\&I into software engineering courses. Based on the same survey data, the paper explores how educators perceive D\&I, identifies implementation barriers, and offers curriculum-level solutions. 

Gama et al. \cite{gama2024s}, describe a graduate-level course that addresses various diversity dimensions, including race, disability, and neurodivergence. The course uses both user-centered and team-centered perspectives, particularly in the context of software requirements and human aspects of software engineering. Teamwork is a fundamental component, with students collaborating remotely through digital platforms. The main results of the paper regard the students' awareness in D\&I themes and the perceived safety during teamworks, with no negative results gathered. 

Murphy et al. \cite{murphy2021incorporating} explored how reading assignments on D\&I topics were integrated into a technically focused undergraduate course. Spanning three semesters and involving nearly 600 students, the intervention aimed to increase awareness of diverse identities and the importance of inclusivity in technology development. The study focuses on individual reflection and class discussions rather than structured teamwork or agile practices.

\section{Conclusions and future work}\label{conclusions}

We analyzed 51 teams distributed over three academic years, measuring their
diversity -- Gender, Nationality and their co-presence, as well as career diversity --
through three distinct indexes.

Through our preliminary analysis of the distinct diversity indexes frequently adopted in the literature, we observe a generally good correlation among the three considered team diversity indexes though a case of inconsistency emerged, thus this is an aspect that deserves further inquiry. 

We observe a mild positive correlation between Gender diversity with team
performance. All other individual diversity indexes have negligible correlations with the team performance. 

Considering the combined effects of various diversity aspects we notice
that team gender and country diversity may play a positive effect on
performance while their co-presence in a team, as well as Career diversities, play
a negative --but not statistically significant-- effect. 

Based on the statistical results and on the analysis of the outcome of the course, as actionable points for educators, we suggest:

\begin{itemize}
\item Encouraging the formation of teams with high diversity for different aspects, since inter-team diversity has no statistically significant negative effect on the outcomes of the team projects.
\item If possible, allocating the teams to the group in a way finalized to maximize diversity, since an autonomous composition of the teams by the students themselves tends to minimize it.
\item Maintaining an active line of support in case intra-group communication issues arise.
    
\end{itemize}

The study detailed in this paper was based solely on quantitative analyses of the results obtained by the teams at the end of the course. Starting from the 2024-2025 edition of the course, we plan to gather -- through closed- or open-ended questionnaires administered to the students -- qualitative and subjective opinions about their experience, and the overall safety and comfort perceived during their activities in their team.

\bibliographystyle{ACM-Reference-Format}
\bibliography{biblio.bib}


\begin{thebibliography}{26}


\ifx \showCODEN    \undefined \def \showCODEN     #1{\unskip}     \fi
\ifx \showISBNx    \undefined \def \showISBNx     #1{\unskip}     \fi
\ifx \showISBNxiii \undefined \def \showISBNxiii  #1{\unskip}     \fi
\ifx \showISSN     \undefined \def \showISSN      #1{\unskip}     \fi
\ifx \showLCCN     \undefined \def \showLCCN      #1{\unskip}     \fi
\ifx \shownote     \undefined \def \shownote      #1{#1}          \fi
\ifx \showarticletitle \undefined \def \showarticletitle #1{#1}   \fi
\ifx \showURL      \undefined \def \showURL       {\relax}        \fi
\providecommand\bibfield[2]{#2}
\providecommand\bibinfo[2]{#2}
\providecommand\natexlab[1]{#1}
\providecommand\showeprint[2][]{arXiv:#2}

\bibitem[{American Psychological Association and others}(2015)]%
        {american2015guidelines}
\bibfield{author}{\bibinfo{person}{{American Psychological Association and
  others}}.} \bibinfo{year}{2015}\natexlab{}.
\newblock \showarticletitle{Guidelines for psychological practice with
  transgender and gender nonconforming people}.
\newblock \bibinfo{journal}{\emph{American psychologist}} \bibinfo{volume}{70},
  \bibinfo{number}{9} (\bibinfo{year}{2015}), \bibinfo{pages}{832--864}.
\newblock


\bibitem[Anderson et~al\mbox{.}(2024)]%
        {anderson2024learning}
\bibfield{author}{\bibinfo{person}{Neil Anderson}, \bibinfo{person}{Aidan
  McGowan}, \bibinfo{person}{Leo Galway}, \bibinfo{person}{Matthew Collins},
  {and} \bibinfo{person}{Philip Hanna}.} \bibinfo{year}{2024}\natexlab{}.
\newblock \showarticletitle{Learning to Improve Gender Equality: An Analysis of
  Software Engineering Education in a {UK} University}. In
  \bibinfo{booktitle}{\emph{Proceedings of the 5th ACM/IEEE Workshop on Gender
  Equality, Diversity, and Inclusion in Software Engineering}}.
  \bibinfo{pages}{37--44}.
\newblock


\bibitem[Blau(1977)]%
        {blau1977}
\bibfield{author}{\bibinfo{person}{Peter~M. Blau}.}
  \bibinfo{year}{1977}\natexlab{}.
\newblock \bibinfo{booktitle}{\emph{Inequality and heterogeneity: A primitive
  theory of social structure}}.
\newblock \bibinfo{publisher}{Free Press}.
\newblock


\bibitem[Chakravorti(2020)]%
        {Chakravorti20}
\bibfield{author}{\bibinfo{person}{Bhaskar Chakravorti}.}
  \bibinfo{year}{2020}\natexlab{}.
\newblock \showarticletitle{To Increase Diversity, {U.S.} Tech Companies Need
  to Follow the Talent}.
\newblock \bibinfo{journal}{\emph{Harvard Business Review}}
  (\bibinfo{year}{2020}).
\newblock
\urldef\tempurl%
\url{https://hbr.org/2020/12/to-increase-diversity-u-s-tech-companies-need-to-follow-the-talent}
\showURL{%
\tempurl}


\bibitem[Crenshaw(2013)]%
        {crenshaw2013demarginalizing}
\bibfield{author}{\bibinfo{person}{Kimberl{\'e} Crenshaw}.}
  \bibinfo{year}{2013}\natexlab{}.
\newblock \showarticletitle{Demarginalizing the intersection of race and sex: A
  black feminist critique of antidiscrimination doctrine, feminist theory and
  antiracist politics}.
\newblock In \bibinfo{booktitle}{\emph{Feminist legal theories}}.
  \bibinfo{publisher}{Routledge}, \bibinfo{pages}{23--51}.
\newblock


\bibitem[Cutrupi et~al\mbox{.}(2024)]%
        {cutrupi2024draw}
\bibfield{author}{\bibinfo{person}{Claudia~Maria Cutrupi},
  \bibinfo{person}{Irene Zanardi}, {and} \bibinfo{person}{Letizia Jaccheri}.}
  \bibinfo{year}{2024}\natexlab{}.
\newblock \showarticletitle{Draw a Software Engineer Test-Preliminary Attempts
  to Investigate University Students’ Perceptions of Software Engineering
  Professions}. In \bibinfo{booktitle}{\emph{Proceedings of the 5th ACM/IEEE
  Workshop on Gender Equality, Diversity, and Inclusion in Software
  Engineering}}. \bibinfo{pages}{45--46}.
\newblock


\bibitem[Gama and Santos(2024)]%
        {gama2024s}
\bibfield{author}{\bibinfo{person}{Kiev Gama} {and} \bibinfo{person}{Reydne
  Santos}.} \bibinfo{year}{2024}\natexlab{}.
\newblock \showarticletitle{It’s not all about gender: A Multi-dimensional
  Course Perspective on Diversity and Inclusion in Software Engineering
  Education}. In \bibinfo{booktitle}{\emph{Simp{\'o}sio Brasileiro de
  Engenharia de Software (SBES)}}. SBC, \bibinfo{pages}{487--498}.
\newblock


\bibitem[Guzmán et~al\mbox{.}(2023)]%
        {Guzmn2023}
\bibfield{author}{\bibinfo{person}{Emitzá Guzmán}, \bibinfo{person}{Ricarda
  Anna-Lena Fischer}, {and} \bibinfo{person}{Janey Kok}.}
  \bibinfo{year}{2023}\natexlab{}.
\newblock \showarticletitle{Mind the gap: gender, micro-inequities and barriers
  in software development}.
\newblock \bibinfo{journal}{\emph{Empirical Software Engineering}}
  \bibinfo{volume}{29}, \bibinfo{number}{1} (\bibinfo{date}{Dec.}
  \bibinfo{year}{2023}).
\newblock
\showISSN{1573-7616}
\href{https://doi.org/10.1007/s10664-023-10379-8}{doi:\nolinkurl{10.1007/s10664-023-10379-8}}


\bibitem[Happe et~al\mbox{.}(2024)]%
        {happe2024decoding}
\bibfield{author}{\bibinfo{person}{Lucia Happe}, \bibinfo{person}{Kai
  Marquardt}, \bibinfo{person}{Ricarda Trumpf}, {and} \bibinfo{person}{Ingo
  Wagner}.} \bibinfo{year}{2024}\natexlab{}.
\newblock \showarticletitle{Decoding the Gap: A Retrospective Analysis of
  Women's Experiences in Software Engineering}. In
  \bibinfo{booktitle}{\emph{Proceedings of the 5th ACM/IEEE Workshop on Gender
  Equality, Diversity, and Inclusion in Software Engineering}}.
  \bibinfo{pages}{27--28}.
\newblock


\bibitem[Hyrynsalmi(2023)]%
        {hyrynsalmi2023diversity}
\bibfield{author}{\bibinfo{person}{Sonja~M Hyrynsalmi}.}
  \bibinfo{year}{2023}\natexlab{}.
\newblock \showarticletitle{How Diversity and Inclusion Are Approached in
  Software Engineering University-level Teaching}. In
  \bibinfo{booktitle}{\emph{2023 IEEE/ACM 4th Workshop on Gender Equity,
  Diversity, and Inclusion in Software Engineering (GEICSE)}}. IEEE,
  \bibinfo{pages}{17--24}.
\newblock


\bibitem[Hyrynsalmi(2024)]%
        {hyrynsalmi2024challenges}
\bibfield{author}{\bibinfo{person}{Sonja~M Hyrynsalmi}.}
  \bibinfo{year}{2024}\natexlab{}.
\newblock \showarticletitle{Challenges and opportunities: Implementing
  diversity and inclusion in software engineering university level education in
  Finland}.
\newblock \bibinfo{journal}{\emph{Journal of Systems and Software}}
  (\bibinfo{year}{2024}), \bibinfo{pages}{112239}.
\newblock


\bibitem[Hyrynsalmi et~al\mbox{.}(2024)]%
        {hyrynsalmi2024second}
\bibfield{author}{\bibinfo{person}{Sonja~M Hyrynsalmi}, \bibinfo{person}{Ella
  Peltonen}, \bibinfo{person}{Fanny Vainionp{\"a}{\"a}}, {and}
  \bibinfo{person}{Sami Hyrynsalmi}.} \bibinfo{year}{2024}\natexlab{}.
\newblock \showarticletitle{The Second Round: Diverse Paths Towards Software
  Engineering}. In \bibinfo{booktitle}{\emph{Proceedings of the 5th ACM/IEEE
  Workshop on Gender Equality, Diversity, and Inclusion in Software
  Engineering}}. \bibinfo{pages}{29--36}.
\newblock


\bibitem[Kovaleva et~al\mbox{.}(2022)]%
        {kovaleva2022designing}
\bibfield{author}{\bibinfo{person}{Yekaterina Kovaleva}, \bibinfo{person}{Ari
  Happonen}, {and} \bibinfo{person}{Eneli Kindsiko}.}
  \bibinfo{year}{2022}\natexlab{}.
\newblock \showarticletitle{Designing gender-neutral software engineering
  program. stereotypes, social pressure, and current attitudes based on recent
  studies}. In \bibinfo{booktitle}{\emph{Proceedings of the Third Workshop on
  Gender Equality, Diversity, and Inclusion in Software Engineering}}.
  \bibinfo{pages}{43--50}.
\newblock


\bibitem[Murphy et~al\mbox{.}(2021)]%
        {murphy2021incorporating}
\bibfield{author}{\bibinfo{person}{Christian Murphy}, \bibinfo{person}{Anya
  Mushakevich}, {and} \bibinfo{person}{Yunha Park}.}
  \bibinfo{year}{2021}\natexlab{}.
\newblock \showarticletitle{Incorporating readings on diversity and inclusion
  into a traditional software engineering course}. In
  \bibinfo{booktitle}{\emph{2021 Conference on Research in Equitable and
  Sustained Participation in Engineering, Computing, and Technology
  (RESPECT)}}. IEEE, \bibinfo{pages}{1--5}.
\newblock


\bibitem[Nguyen-Duc and Jaccheri(2023)]%
        {NguyenJaccheri}
\bibfield{author}{\bibinfo{person}{Anh Nguyen-Duc} {and}
  \bibinfo{person}{Letizia Jaccheri}.} \bibinfo{year}{2023}\natexlab{}.
\newblock \showarticletitle{Gender Equality in Software Engineering Education
  -- A Study of Female Participation in Customer-Driven Projects}. In
  \bibinfo{booktitle}{\emph{Sustainability in Software Engineering and Business
  Information Management}}, \bibfield{editor}{\bibinfo{person}{Varun Gupta},
  \bibinfo{person}{Luis Rubalcaba}, \bibinfo{person}{Chetna Gupta}, {and}
  \bibinfo{person}{Thomas Hanne}} (Eds.). \bibinfo{publisher}{Springer
  International Publishing}, \bibinfo{address}{Cham}, \bibinfo{pages}{39--49}.
\newblock
\showISBNx{978-3-031-32436-9}


\bibitem[Ortu et~al\mbox{.}(2017)]%
        {Ortu2017}
\bibfield{author}{\bibinfo{person}{Marco Ortu}, \bibinfo{person}{Giuseppe
  Destefanis}, \bibinfo{person}{Steve Counsell}, \bibinfo{person}{Stephen
  Swift}, \bibinfo{person}{Roberto Tonelli}, {and} \bibinfo{person}{Michele
  Marchesi}.} \bibinfo{year}{2017}\natexlab{}.
\newblock \showarticletitle{How diverse is your team? Investigating gender and
  nationality diversity in GitHub teams}.
\newblock \bibinfo{journal}{\emph{Journal of Software Engineering Research and
  Development}} \bibinfo{volume}{5}, \bibinfo{number}{1}
  (\bibinfo{year}{2017}), \bibinfo{pages}{9}.
\newblock
\showISBNx{2195-1721}
\href{https://doi.org/10.1186/s40411-017-0044-y}{doi:\nolinkurl{10.1186/s40411-017-0044-y}}


\bibitem[Petrescu et~al\mbox{.}(2023)]%
        {petrescu2023women}
\bibfield{author}{\bibinfo{person}{Manuela~Andreea Petrescu},
  \bibinfo{person}{Simona Motogna}, {and} \bibinfo{person}{Liviu Berciu}.}
  \bibinfo{year}{2023}\natexlab{}.
\newblock \showarticletitle{Women in Scrum Master Role: Challenges and
  Opportunities}. In \bibinfo{booktitle}{\emph{2023 IEEE/ACM 4th Workshop on
  Gender Equity, Diversity, and Inclusion in Software Engineering (GEICSE)}}.
  IEEE, \bibinfo{pages}{49--55}.
\newblock


\bibitem[{Politecnico di Torino}(2024)]%
        {polito2024}
\bibfield{author}{\bibinfo{person}{{Politecnico di Torino}}.}
  \bibinfo{year}{2024}\natexlab{}.
\newblock \bibinfo{booktitle}{\emph{Diversità è Cambiamento - Bilancio di
  Genere 2023}}.
\newblock
\urldef\tempurl%
\url{{https://www.polito.it/sites/default/files/2024-01/BDG\_2023\_DEF.pdf}}
\showURL{%
\tempurl}


\bibitem[Rodr{\'\i}guez-P{\'e}rez et~al\mbox{.}(2021)]%
        {rodriguez2021perceived}
\bibfield{author}{\bibinfo{person}{Gema Rodr{\'\i}guez-P{\'e}rez},
  \bibinfo{person}{Reza Nadri}, {and} \bibinfo{person}{Meiyappan Nagappan}.}
  \bibinfo{year}{2021}\natexlab{}.
\newblock \showarticletitle{Perceived diversity in software engineering: a
  systematic literature review}.
\newblock \bibinfo{journal}{\emph{Empirical Software Engineering}}
  \bibinfo{volume}{26} (\bibinfo{year}{2021}), \bibinfo{pages}{1--38}.
\newblock


\bibitem[S\'{a}nchez-Gord\'{o}n and Colomo-Palacios(2024)]%
        {Sanchez2024}
\bibfield{author}{\bibinfo{person}{Mary S\'{a}nchez-Gord\'{o}n} {and}
  \bibinfo{person}{Ricardo Colomo-Palacios}.} \bibinfo{year}{2024}\natexlab{}.
\newblock \showarticletitle{On the Intersectionality of Software Practitioners
  and Role Models}. In \bibinfo{booktitle}{\emph{Proceedings of the 5th
  ACM/IEEE Workshop on Gender Equality, Diversity, and Inclusion in Software
  Engineering}} (Lisbon, Portugal) \emph{(\bibinfo{series}{GE@ICSE '24})}.
  \bibinfo{publisher}{Association for Computing Machinery},
  \bibinfo{address}{New York, NY, USA}, \bibinfo{pages}{22–26}.
\newblock
\showISBNx{9798400705755}
\href{https://doi.org/10.1145/3643785.3648483}{doi:\nolinkurl{10.1145/3643785.3648483}}


\bibitem[Spellerberg and Fedor(2003)]%
        {Shannon}
\bibfield{author}{\bibinfo{person}{Ian~F. Spellerberg} {and}
  \bibinfo{person}{Peter~J. Fedor}.} \bibinfo{year}{2003}\natexlab{}.
\newblock \showarticletitle{A tribute to {Claude Shannon} (1916–2001) and a
  plea for more rigorous use of species richness, species diversity and the
  ‘{Shannon–Wiener}’ Index}.
\newblock \bibinfo{journal}{\emph{Global Ecology and Biogeography}}
  \bibinfo{volume}{12}, \bibinfo{number}{3} (\bibinfo{year}{2003}),
  \bibinfo{pages}{177--179}.
\newblock
\href{https://doi.org/10.1046/j.1466-822X.2003.00015.x}{doi:\nolinkurl{10.1046/j.1466-822X.2003.00015.x}}
\showeprint{https://onlinelibrary.wiley.com/doi/pdf/10.1046/j.1466-822X.2003.00015.x}


\bibitem[Stahl et~al\mbox{.}(2010)]%
        {Stahl2010}
\bibfield{author}{\bibinfo{person}{G{\"u}nter~K Stahl},
  \bibinfo{person}{Martha~L Maznevski}, \bibinfo{person}{Andreas Voigt}, {and}
  \bibinfo{person}{Karsten Jonsen}.} \bibinfo{year}{2010}\natexlab{}.
\newblock \showarticletitle{Unraveling the effects of cultural diversity in
  teams: A meta-analysis of research on multicultural work groups}.
\newblock \bibinfo{journal}{\emph{Journal of International Business Studies}}
  \bibinfo{volume}{41}, \bibinfo{number}{4} (\bibinfo{year}{2010}),
  \bibinfo{pages}{690--709}.
\newblock
\showISBNx{1478-6990}
\href{https://doi.org/10.1057/jibs.2009.85}{doi:\nolinkurl{10.1057/jibs.2009.85}}


\bibitem[Szlavi et~al\mbox{.}(2023)]%
        {Szlavi23}
\bibfield{author}{\bibinfo{person}{Anna Szlavi},
  \bibinfo{person}{Marit~Fredrikke Hansen}, \bibinfo{person}{Sandra~Helen
  Husnes}, {and} \bibinfo{person}{Tayana~Uchôa Conte}.}
  \bibinfo{year}{2023}\natexlab{}.
\newblock \showarticletitle{Intersectionality in Computer Science: A Systematic
  Literature Review}. In \bibinfo{booktitle}{\emph{2023 IEEE/ACM 4th Workshop
  on Gender Equity, Diversity, and Inclusion in Software Engineering
  (GEICSE)}}. \bibinfo{pages}{9--16}.
\newblock
\href{https://doi.org/10.1109/GEICSE59319.2023.00006}{doi:\nolinkurl{10.1109/GEICSE59319.2023.00006}}


\bibitem[Vasilescu et~al\mbox{.}(2015)]%
        {vasilescu2015gender}
\bibfield{author}{\bibinfo{person}{Bogdan Vasilescu}, \bibinfo{person}{Daryl
  Posnett}, \bibinfo{person}{Baishakhi Ray}, \bibinfo{person}{Mark~GJ van~den
  Brand}, \bibinfo{person}{Alexander Serebrenik}, \bibinfo{person}{Premkumar
  Devanbu}, {and} \bibinfo{person}{Vladimir Filkov}.}
  \bibinfo{year}{2015}\natexlab{}.
\newblock \showarticletitle{Gender and tenure diversity in {GitHub} teams}. In
  \bibinfo{booktitle}{\emph{Proceedings of the 33rd annual ACM conference on
  human factors in computing systems}}. \bibinfo{pages}{3789--3798}.
\newblock


\bibitem[Wu(2012)]%
        {wu2012overview}
\bibfield{author}{\bibinfo{person}{Shujian Wu}.}
  \bibinfo{year}{2012}\natexlab{}.
\newblock \showarticletitle{Overview of communication in global software
  development process}. In \bibinfo{booktitle}{\emph{Proceedings of 2012 IEEE
  International Conference on Service Operations and Logistics, and
  Informatics}}. IEEE, \bibinfo{pages}{474--478}.
\newblock


\bibitem[Zhao and Young(2023)]%
        {zhao2023workplace}
\bibfield{author}{\bibinfo{person}{Xin Zhao} {and} \bibinfo{person}{Riley
  Young}.} \bibinfo{year}{2023}\natexlab{}.
\newblock \showarticletitle{Workplace Discrimination in Software Engineering:
  Where We Stand Today}. In \bibinfo{booktitle}{\emph{2023 IEEE/ACM 45th
  International Conference on Software Engineering: Software Engineering in
  Society (ICSE-SEIS)}}. IEEE, \bibinfo{pages}{188--193}.
\newblock


\end{thebibliography}


\end{document}